\documentclass[aps,prb,nolongbibliography,notitlepage,twocolumn,noeprint,superscriptaddress]{revtex4-2}

\usepackage{amsmath,amssymb,wasysym}
\usepackage{graphicx}
\usepackage{dcolumn}
\usepackage{bm}
\usepackage{braket}
\usepackage{verbatim}
\usepackage{float}
\usepackage{bbold}
\usepackage{color}
\usepackage{xcolor}
\usepackage{relsize}
\usepackage{amsthm}
\usepackage{enumerate}
\usepackage{siunitx}
\usepackage{soul,xcolor}
\usepackage[T1]{fontenc}
\usepackage[colorlinks=true ,urlcolor=blue,urlbordercolor={0 1 1}]{hyperref}

\newcommand{\beq}{\begin{eqnarray}}
\newcommand{\eeq}{\end{eqnarray}}

\newcommand{\bsp}{\begin{split}}
\newcommand{\esp}{\end{split}}

\newcommand{\be}{\begin{equation}}
\newcommand{\ee}{\end{equation}}

\graphicspath{{./figures/}}
\def\bea{\begin{eqnarray}}
\def\eea{\end{eqnarray}}

\makeatletter

\def\env@sqcases{%
  \let\@ifnextchar\new@ifnextchar
  \left\lbrack
  \def\arraystretch{1.2}%
  \array{@{}l@{\quad}l@{}}%
}
\makeatother

\usepackage{mathtools} 
\begin{document}

\title{From Ferrimagnetic Insulator to superconducting Luther-Emery Liquid:\\ A DMRG Study of the Two-Leg Lieb Lattice}

\author{Alexander Nikolaenko}
\affiliation{Department of Physics, Harvard University, Cambridge MA 02138, USA}

\author{Subir Sachdev}
\affiliation{Department of Physics, Harvard University, Cambridge MA 02138, USA}
\affiliation{Center for Computational Quantum Physics, Flatiron Institute, 162 5th Avenue, New York, NY 10010, USA}

\begin{abstract}
Motivated by recent experiments on ultracold, fermionic, spin-1/2 $^6$Li atoms in the Lieb lattice at various Hubbard repulsion $U$ and filling fractions $n$ (Lebrat {\it et al.\/} \href{https://arxiv.org/abs/2404.17555}{arXiv:2404.17555}), we conduct a density matrix renormalization group (DMRG) analysis of the Hubbard model on a two-leg Lieb ladder. At half-filling, we find a ferrimagnetic Mott insulating ground state, consistent with Lieb’s theorem. Away from half-filling, the state with finite total spin $\vec{S}^2 \neq0$, and zero charge gap, persists down to filling $n_c\approx2/3$. For lower, incommensurate fillings, the system is described by a Luttinger liquid with one charge and one spin mode. Intriguingly,  in the small window close to the onset of ferromagnetic order at $n_c=2/3$, we identify a superconducting Luther-Emery phase, suggesting $s_{xy}-$wave pairing.

\end{abstract}
\pacs{Valid PACS appear here}
\maketitle{}
\section{Introduction}

The Hubbard model is one of the simplest yet most powerful models in condensed matter physics, capturing a wide range of strongly correlated phenomena. Under the Hubbard interaction, a noninteracting metal could develop charge, spin or superconducting orders, undergo a Mott transition, or even give rise to exotic fractionalized states such as spin liquid. There has been an extensive effort to understand the ground state of the two dimensional Hubbard model over the last 40 years, see Ref. \cite{Arovas2022,Qin2022} for recent reviews.

\begin{figure}[b!]
\begin{minipage}[h]{1\linewidth}
  \center{\includegraphics[width=0.9\linewidth]{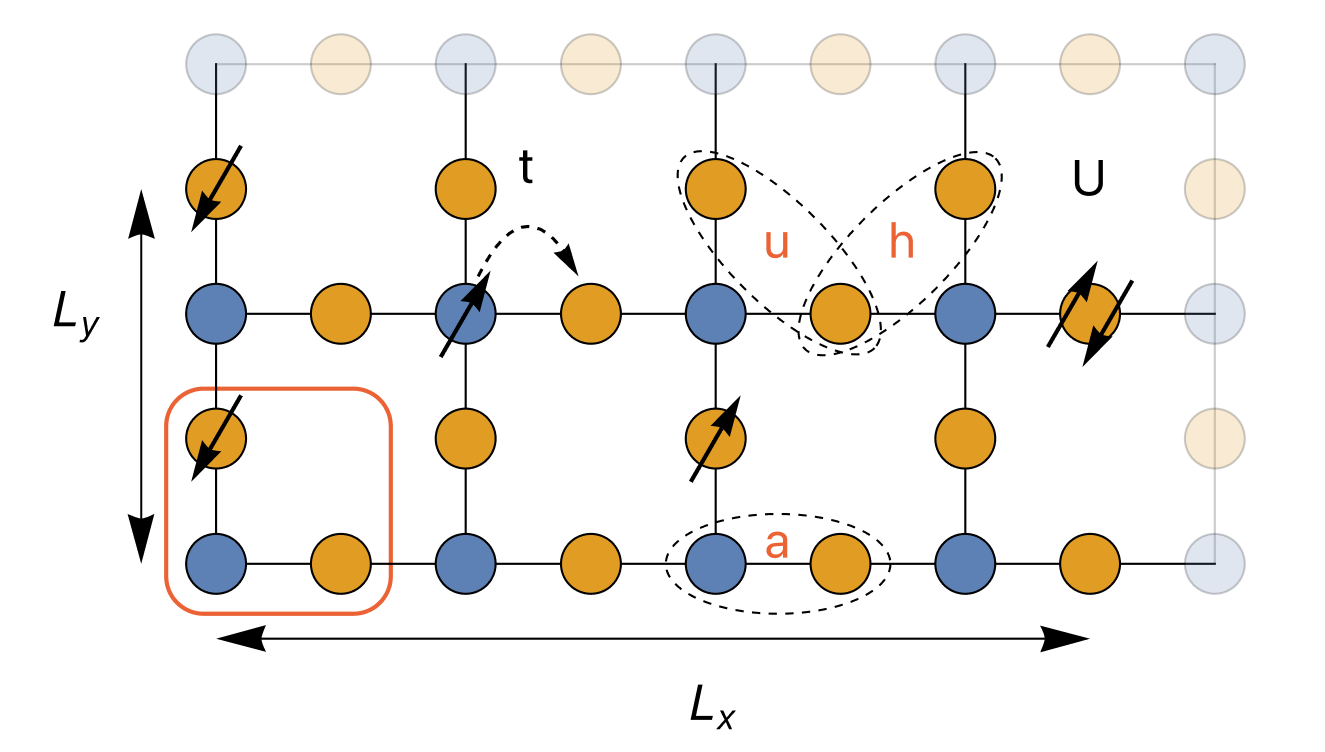}}
        \caption{\label{fig:geometry} Geometry and boundary conditions of the Hubbard model on a Lieb lattice. A unit cell contains three sites as shown by red rounded square. Different bonds are shown by the dashed black lines. The lattice dimensions are $L_x = 4$ and $L_y = 2$ unit cells. Periodic boundary conditions (PBC) are applied in the $y$-direction, indicated by the faded top row which is equivalent to the bottom row.}
    \end{minipage}
    
\end{figure}

While early studies focused on the Hubbard model on the square lattice, recent attempts have been made to understand the fascinating physics of the Hubbard model on other geometries, such as the triangular lattice~\cite {Szasz2020}, the Kagome lattice~\cite {Liao2017}. In this paper, we study the Hubbard model on the Lieb lattice - a two-dimensional bipartite lattice that can be constructed by removing every fourth site from a square lattice, see Fig.~\ref{fig:geometry}.

The Lieb lattice has attracted considerable interest within the scientific community, as it provides a valuable starting point for describing high-temperature cuprate superconductors. The key physics in those materials arises from the CuO$_2$ planes, where the copper atoms are arranged in the square lattice, while the oxygen atoms sit on the bonds, forming the Lieb lattice pattern. The minimal tight-binding model for cuprates involves overlapping Cu $d_{x^2-y^2}$ orbitals and O $p_x,p_y$ orbitals, as was originally proposed by Emery~\cite{Emery1987}. The Emery model has been extensively studied in the context of cuprate  physics~\cite{Jiang2023,Jiang2023_2,Yang2024} using the DMRG approach. Recent advances in ultracold atom experiments have also opened a pathway toward realizing and simulating such multi-orbital lattice models in optical lattices~\cite{lange2026}. These studies primarily focused on the parameter regime relevant to cuprate superconductors, including a finite charge transfer gap between $p$ and $d$ sites as well as distinct hopping amplitudes and Hubbard interactions on the two sublattices. 

Beyond its relevance to cuprate superconductors, the model has opened a direct pathway to the physics of strongly interacting electrons. It was originally explored by Lieb~\cite{Lieb1989}, who rigorously showed that the ground state of any bipartite lattice at half-filling with positive Hubbard interaction strength $U>0$ has a nonzero total spin $s=(|B|-|A|)/2$, where $|A|,|B|$ denote the number of sites in the two sublattices. The Lieb lattice has $|B|=2|A|$ so the ground state at half-filling is ferrimagnetic: it has antiferromagnetic local ordering and nonzero total spin. Beyond conventional ferromagnetic and antiferromagnetic orders, there has been growing interest in more exotic magnetic states such as altermagnetism~\cite{Smejkal2022}, exhibiting vanishing net magnetization and momentum space spin splitting. Such an order can be realized on the Lieb lattice with minimal modifications~\cite{Antonenko2025}.

 \begin{figure*}[t!]
\begin{minipage}[h]{1\linewidth}
  \center{\includegraphics[width=1\linewidth]{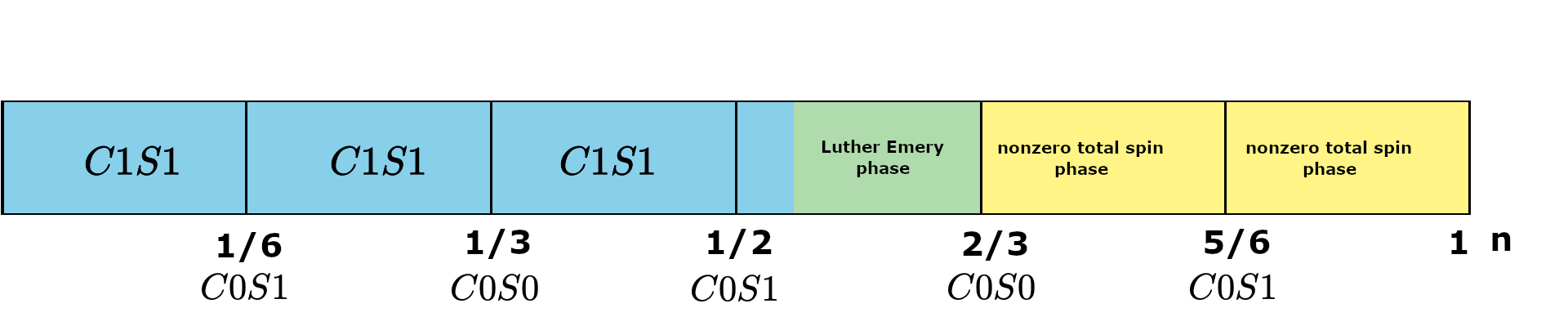}}
  \end{minipage} 
\caption{\label{fig:phase_diagram}Phase diagram of the model as a function of filling $n$ for the Hubbard interaction $U=16$.}
\label{fig:U_16_phase_diagram}
\end{figure*}
The magnetic instabilities of the Lieb lattice at half-filling are closely connected to the presence of a flat band in its tight-binding spectrum at $U=0$. The suppression of kinetic energy enhances the role of interactions, leading to strongly correlated behavior. Thus, flat-band models have become central to the study of magnetism, unconventional superconductivity, and topological phases~\cite{Leykam2018,Kopnin2011,Bergholtz2013}. Recently, it has become possible to realize flat-band models in moiré materials such as twisted bilayer graphene. In these systems, nearly flat moiré bands host a variety of unconventional phases~\cite{Cao2018,Balents2020,Lu2025}.

Beyond the flat-band limit, it is also important to consider the behavior of the system in the vicinity of a magnetic quantum phase transition. In strongly correlated systems, the region near a transition between magnetically ordered and paramagnetic phases is often characterized by enhanced fluctuations, which can give rise to competing and intertwined orders. In particular, critical fluctuations near such transitions can mediate effective attractive interactions between electrons, potentially leading to enhanced superconductivity~\cite{Scalapino2012,Abanov2020}.

A significant step forward has been made in realizing flat-band electronic models using ultracold atoms in optical lattices~\cite{Taie2015,Jo2012}. In particular, a recent experiment by Lebrat {\it et al.\/}\cite{Lebrat24} showed the signatures of the ferrimagnetic state at half-filling in a multi-band Hubbard model in an optical Lieb lattice, using fermionic lithium-6 atoms. They further shed light on what happens away from half-filling, reporting  a doubling of the compressibility
curve at strong interactions.

 Motivated by the experimental findings and supported by theoretical insights, we performed a Hartree-Fock analysis of the Hubbard model on the Lieb lattice at quarter filling~\cite{Nikolaenko2025}. We find that at $U>U_c\approx 4$ the system develops magnetic order: initially forming a spiral state, then evolving into a canted phase with increasing 
$U$, and ultimately becoming fully ferromagnetic at large $U$. This magnetically polarized state naturally accounts for the doubling in compressibility observed in experiments. We further explored alternative routes for explaining the experimental data, including possible electron fractionalization and the emergence of a $\mathbb{Z}_2$ spin-liquid phase.

 While Hartree-Fock calculations could be performed for large lattice sizes, they often overestimate ordering tendencies, and typically perform poorly for strongly-interacting systems, particularly in the vicinity of quantum critical points where fluctuations become dominant. Therefore, in this work we will try to understand the physics of Hubbard model on the two-leg Lieb lattice using the DMRG~\cite{Tenpy2018} method. 

The paper is organized as follows: in Section~\ref{sec:model}, we introduce the model and the geometry; in Section~\ref{sec:phase_diagram}, we describe the phase diagram at strong Hubbard interaction as a function of filling and explain it using the weak coupling limit. Finally, in Section~\ref{sec:superconducting}, we discuss the emerging superconducting Luther Emery phase, analyzing correlation lengths and pairing symmetries as well as its stability at weak coupling. We conclude in Section~\ref{sec:conclusion} and provide technical details in Appendix~\ref{app:details}.

\begin{figure}[b!]
\begin{minipage}[h]{1\linewidth}
  \center{\includegraphics[width=0.8\linewidth]{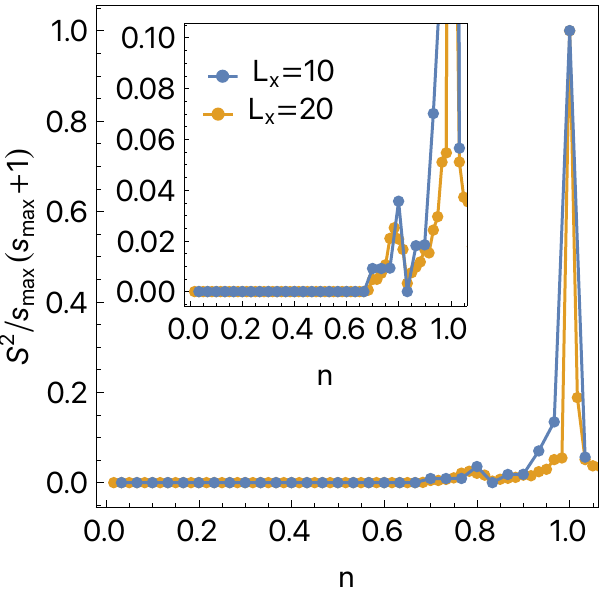}}
        \caption{\label{fig:Stot} Total spin $S^2$ normalized by $s_{\text{max}}(s_{\text{max}}+1)$, where $s_{\text{max}} = L_x L_y / 2$. The blue and yellow curves correspond to $L_x=10$ and $L_x=20$, respectively, calculated at a bond dimension of $\chi=2000$ and $U=16$. The inset shows a magnified view of the same data.}
    \end{minipage}
\end{figure}

\section{Model}
\label{sec:model}

We study the Hubbard model on a two-leg Lieb ladder shown in Fig.~\ref{fig:geometry}. The Hamiltonian reads
\begin{equation}
H = -t \sum_{\langle i,j \rangle, \sigma} \left( c_{i\sigma}^\dagger c_{j\sigma} + \text{h.c.} \right) + U \sum_{i} n_{i\uparrow} n_{i\downarrow} 
\end{equation}

 \begin{figure*}[t]
\begin{minipage}[h]{1\linewidth}
  \center{\includegraphics[width=1\linewidth]{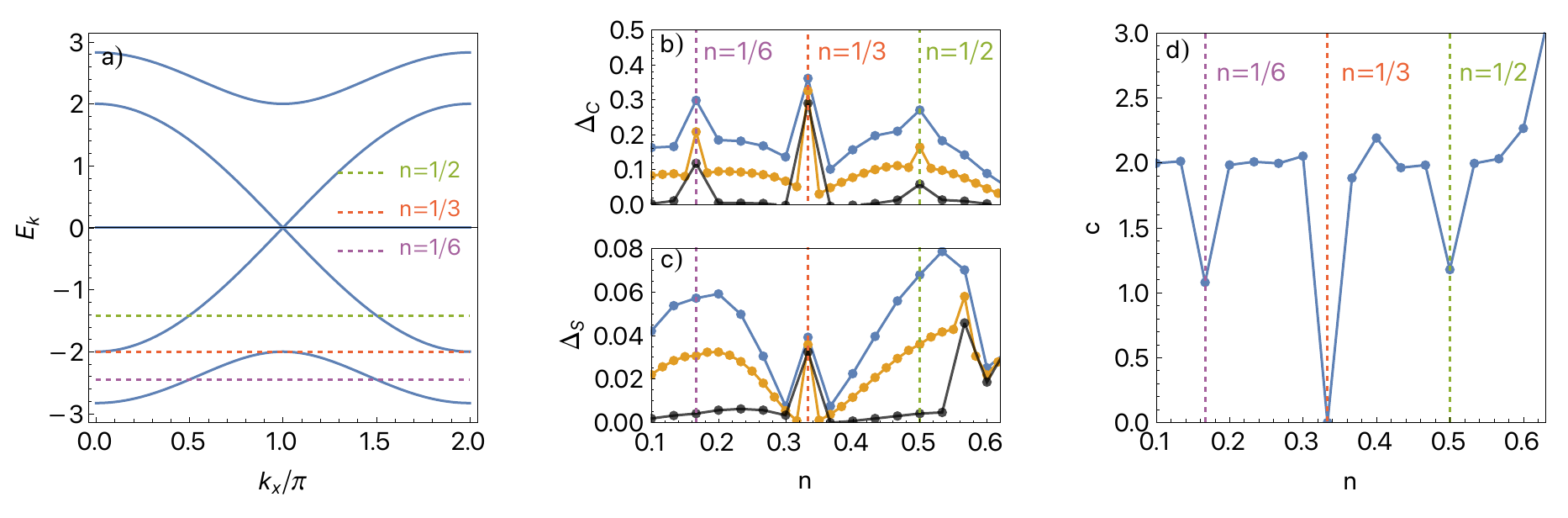}}
  \end{minipage} 
\caption{(a) The dispersion of 2-ladder Lieb lattice with $U=0$. Green, red, and purple dashed lines correspond to commensurate fillings $n=1/2$, $1/3$, and $1/6$, respectively. Charge (b) and spin (c) gaps as a function of filling $n$ at $U=16$. The blue, orange, and black lines correspond to  $L_x=10$, $L_x=20$ and extrapolation to  $L_x \rightarrow \infty$. (d) Central charge as a function of filling at $U=16$ extracted via iDMRG on with a unit cell of length $L_x=10$. Maximum bond dimension used is $\chi=3000$. }
\label{fig:2lad}
\end{figure*}

Here $c_{i\sigma}^\dagger$ creates a fermion with spin $\sigma = \uparrow, \downarrow$ on site $i$, and $n_{i\sigma} = c_{i\sigma}^\dagger c_{i\sigma}$. The hopping amplitude $t$ connects nearest neighbors of the Lieb lattice geometry, and $U > 0$ denotes the on-site repulsion.

The unit cell consists of three sites, shown by the red rounded square in Fig.~\ref{fig:geometry}. The ladder has $L_x$ unit cells in the $x$-direction and $L_y$ unit cells in the $y$-direction. Due to exponentially growing complexity with $L_y$, we restrict ourselves to the two-leg Lieb ladder with $L_y=2$. The total number of sites is $N_{tot}=3 L_x L_y$. We define the filling as $n=N_e/N_{tot}$, where $N_e$ is the number of electrons; thus half-filling corresponds to $n=1$.

We impose periodic boundary conditions (PBC) along the $y$-direction. For finite DMRG we use open boundary conditions along $x$-direction, on systems up to $L_x=30$ with bond dimensions up to $\chi=4000$. For  infinite DMRG (iDMRG), we use periodic boundary conditions (PBC) along $x$-direction with a unit cell up to $L_x=10$ and bond dimensions up to $\chi=8000$.

\section{Phase diagram}
\label{sec:phase_diagram}

In this section, we determine the ground-state phase diagram of the two-leg Lieb ladder as a function of filling $n$ at strong interaction $U=16$. Before discussing each phase in more detail, we provide a brief overview of the phase diagram shown in Fig.~\ref{fig:phase_diagram}. 

At filling $n=1$, we find a ferrimagnetic insulator consistent with Lieb's original prediction~\cite{Lieb1989}. The total spin remains non-zero in the region $n \in (2/3, 1)$ until it vanishes at smaller $n$. In the small region close to the ferrimagnetic phase ($n \in (0.55, 2/3)$), we find a superconducting Luther-Emery phase, which will be the primary focus of our subsequent analysis. The remainder of the phase diagram for fillings $n \in (0, 0.55)$ consists of a paramagnetic phase, characterized as a Luttinger liquid with one charge and one spin mode (C1S1). Finally, at the commensurate fillings $n=1/6$ and $1/2$, we find a Mott insulator with one spin mode (C0S1), while at $n=1/3$, we find a Mott insulator with a spin gap (C0S0). We now turn to a detailed discussion of each of these phases.

\subsection{$S^2_{tot}\neq 0$ phase}
Lieb showed~\cite{Lieb1989} that at half-filling, the system has a total nonzero spin $S_{\text{max}}^2=s_{\text{max}}(s_{\text{max}}+1)$ with $s_{\text{max}}=L_x L_y/2$ for arbitrary Hubbard interaction strength $U$. However, there are no rigorous statements away from half-filling, and the fate of the ferromagnetism is generally unclear. Fig.~\ref{fig:Stot} shows the total spin, normalized by the value $S_{\text{max}}^2$ at half-filling. As expected, at $n=1$, the total spin coincides with the value predicted by Lieb. Away from half-filling, the total spin quickly decreases and becomes on the order of $S^2/S_{\text{max}}^2 \lesssim 0.1$. Ferromagnetic order is detected for all fillings above $n > n_c$, where $n_c=2/3$. The presence of ferromagnetism is closely linked to the flat bands observed in the non-interacting system at these fillings. In this regime, DMRG convergence becomes challenging, and large bond dimensions are necessary to achieve accurate results.

\subsection{Commensurate fillings}

To explain the rest of the phase diagram at $n<2/3$ as the system enters the paramagnetic region, it is instructive to start with the
non-interacting limit $U=0$. In this case, the tight-binding spectrum can be found exactly. It has two bonding and antibonding bands $E_0(k_x)=\pm \sqrt{2t(3+\cos k_x)}$, $E_\pi(k_x)=\pm \sqrt{2t(1+\cos k_x)}$ as well as two flat bands at $E(k_x)=0$~\cite{Nikolaenko2025,Jonah2025}.

Fig.~\ref{fig:2lad}(a) shows the dispersion of the 2-leg Lieb lattice in momentum space. In the incommensurate case, the system has only one pair of Fermi points and the resulting phase is the C1S1 Luttinger liquid with one charge and one spin mode.

At commensurate fillings $n=1/2$ and $n=1/6$ the chemical potential lies in the middle of the $E_\pi(k_x)$ and $E_0(k_x)$ bands, correspondingly, see Fig.~\ref{fig:2lad}(a).  In this regime, the momentum $k_F=\pi/2$ is commensurate  with the reciprocal lattice wavevector and the Umklapp processes at $4k_F$ are allowed~\cite{Giamarchi2003}. Once the interaction is introduced, the phase is unstable to developing a charge gap.

At the commensurate filling $n=1/3$, the chemical potential lies in between the filled $E_0(k_x)$ and empty $E_\pi(k_x)$ band. The particle-hole excitations carry the momenta $k_x=\pi/2$ and have quadratic dispersion $E\sim \delta  k_x^2$. The phase is not described by a Luttinger liquid but instead is a quantum critical point with $z=2$~\cite{Sachdev_2011}. The diverging density of states makes the system highly unstable to interactions. Note, that in two dimensions the quadratic band touching leads to a plethora of interesting phases, such as Quantum Spin Hall, nematic and spin-ordered phases \cite{Sun2009,Herbut2014}. However, in our case, the quadratic band touching is specific only to the 2-leg ladder and disappears once $L_y>2$.

Fig.~\ref{fig:2lad}(b) shows the charge gap as a function of filling once the strong interaction $U=16$ is introduced.  It is defined as $\Delta_c=(E(N_e+2)+E(N_e-2)-2E(N_e))/2$, where $E(N_e)$ is the ground state energy and $N_e$ is the total number of electrons. At the commensurate fillings $n=\mathbb{Z}/6$ the charge gap becomes finite, as expected from the previous analysis at $U\rightarrow0$.

Fig.~\ref{fig:2lad}(c) shows the spin gap $\Delta_S=E(S_z=1)-E(S_z=0)$, where $E(S_z)$ is the ground state energy in the sector with total spin $S_z$. At fillings $n=1/6$ and $n=1/2$ no spin gap develops, as the system is similar to the attractive 1d Hubbard model. At filling $n=1/3$ there is a small, but finite spin gap on the order of $\Delta_S \approx 0.03$. Finally, as the filling approaches the ferromagnetic phase, another finite spin gap develops. It would be discussed extensively in the next section. 

In Fig.~\ref{fig:2lad}(d), we computed the total central charge at different fillings, using iDMRG~\cite{Tenpy2018} on a  periodic ladder with $L_x=10$ unit cells, see Appendix \ref{app:details} for technical details. As expected, at incommensurate fillings the phase is C1S1 with central charge $c=2$. At commensurate fillings $n=1/6$ and $n=1/2$ the charge mode becomes gapped and the phase is C0S1 with $c=1$. At $n=1/3$ both charge and spin modes are gapped and the phase is a Mott insulator C0S0 with $c=0$. We also note that as we approach the ferromagnetic critical point $n_c=2/3$ the central charge increased $c>2$, which we attribute to the slow convergence of the entanglement entropy in the presence of the nearly-flat bands.
\begin{figure}[t!]
\begin{minipage}[h]{1\linewidth}
  \center{\includegraphics[width=0.9\linewidth]{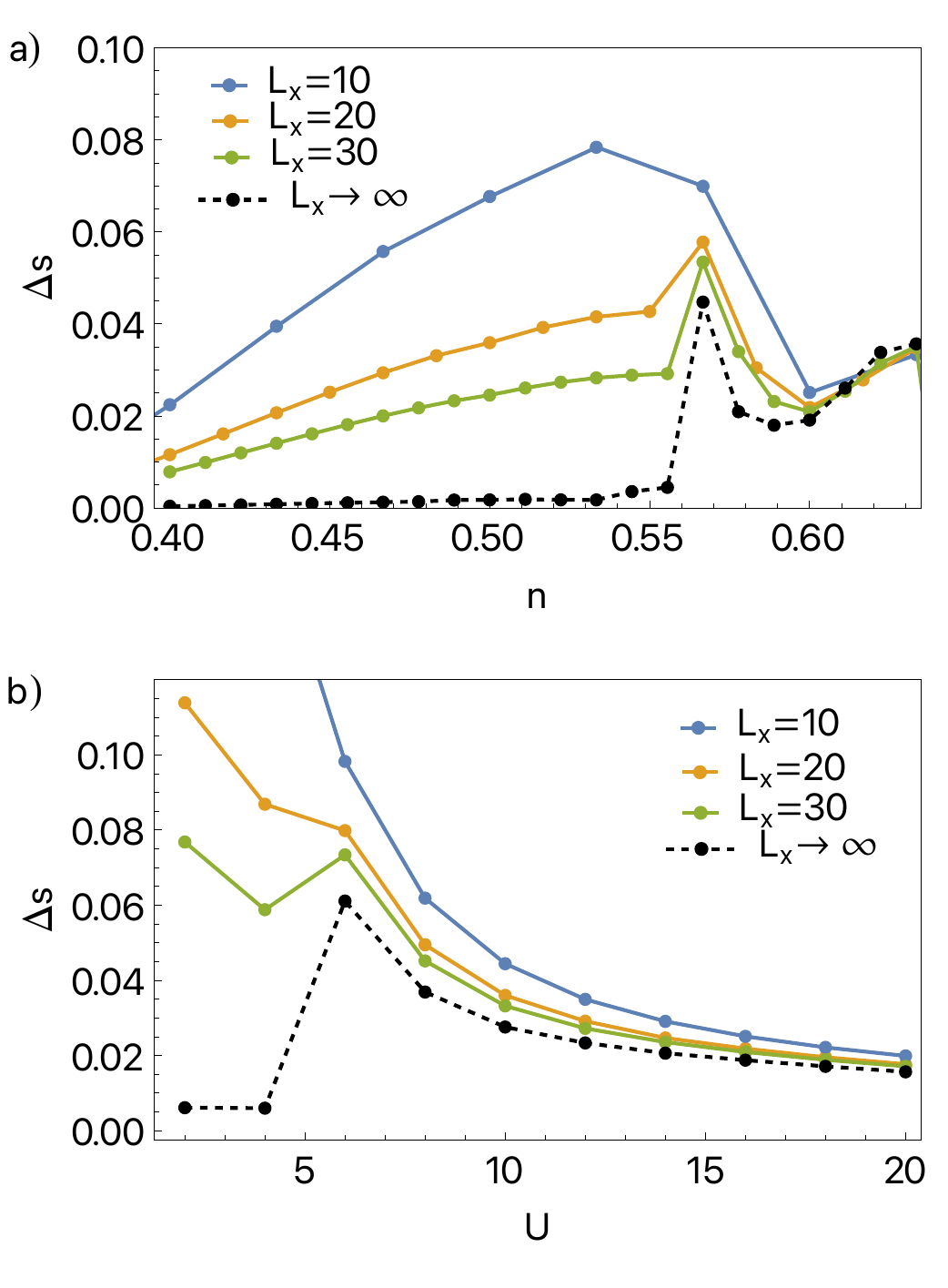}}
        \caption{\label{fig:spin_gap} (a) The spin gap as a function of filling $n$ at fixed interaction strength $U=16$. (b) The spin gap as a function of $U$ at fixed filling $n=0.6$. Results are shown for system sizes $L_x = 10$ (blue), $20$ (yellow), and $30$ (green). The black dashed line represents the extrapolation to $L_x \rightarrow \infty$. Calculations were performed using a maximum bond dimension of $\chi=3000$ to ensure convergence.   }
    \end{minipage}
\end{figure}

\begin{figure*}[t!]
\begin{minipage}[h]{1\linewidth}
  \center{\includegraphics[width=1\linewidth]{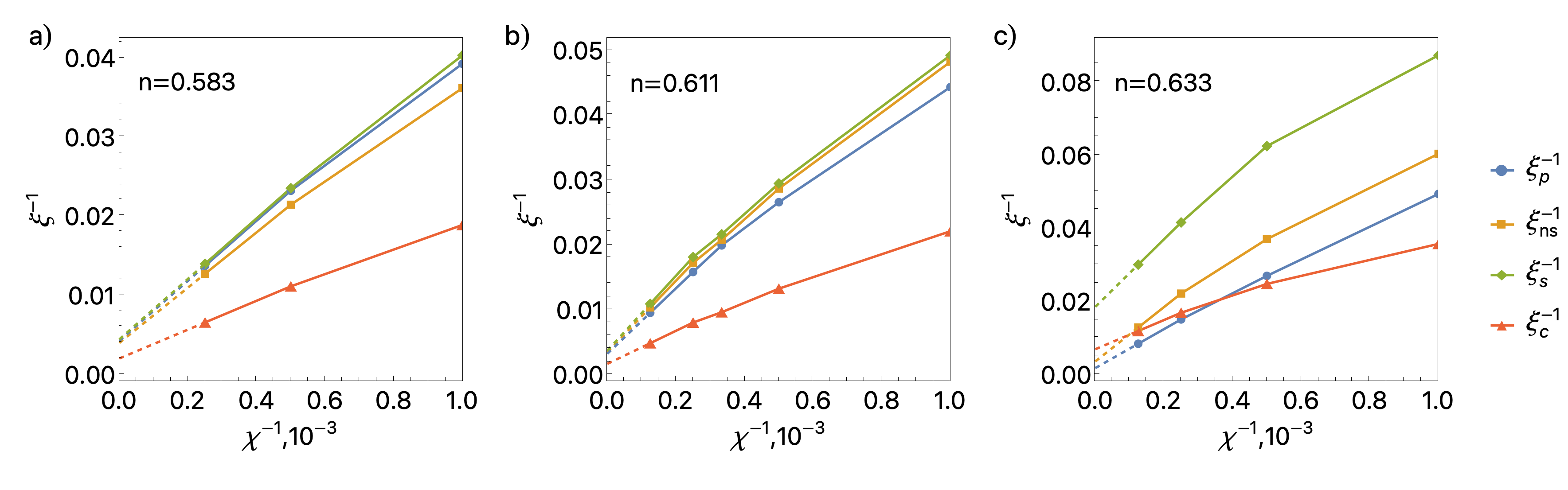}}
        \caption{\label{fig:corr_length} Inverse correlation lengths as a function of inverse bond dimension $\chi^{-1}$ for three different fillings: (a) $n\approx0.583$ with $L_x=8$ and $N_e=28$,
        (b) $n\approx0.611$ with  $L_x=6$ and $N_e=22$, and (c) $n\approx0.633$ with $L_x=10$ and $N_e=38$. The blue, yellow, green, and red lines correspond to the pairing ($\xi_p$), density ($\xi_{ns}$), spin ($\xi_s$), and charge ($\xi_c$) correlation lengths, while the dashed lines correspond to extrapolation to $\chi\rightarrow\infty$. Calculations were performed using iDMRG with a maximum bond dimension of $\chi=8000$ at fixed interaction strength $U=16$. }
    \end{minipage}
\end{figure*}

\section{superconducting phase}
\label{sec:superconducting}
While most of the phase diagram can be understood from the non interacting limit $U \rightarrow0$, the discovery of the spin gap close to the ferromagnetic phase is surprising. 
Fig.~\ref{fig:spin_gap}(a) shows the spin gap as a function of filling for system sizes up to $L_x=30$. In the region $n\in(0.55,2/3)$ there is a clear gap on the order of $\Delta_S\approx0.03$.

Fig.~\ref{fig:spin_gap}(b) shows the spin gap as a function of Hubbard interaction $U$ for filling $n=0.6$ close to the critical point $n_c=2/3$. As we decrease $U$, the spin gap increases, until it rapidly jumps to zero at $U \lesssim6$. We also find that at lower fillings, such as $n=0.56$, a relatively strong interaction ($U \gtrsim 10$) is required to open a finite spin gap. These results indicate that the spin gap is intrinsically linked to the strong-interaction regime in the vicinity of the quantum critical point (QCP).

To understand whether the superconducting or density fluctuations dominate inside the superconducting phase, one has to analyze the decay of the correlation functions. We found that the in finite DMRG the boundary effects are quite strong, and large system sizes are needed to reach definitive conclusions. Therefore, we directly analyzed the correlation lengths extracted from iDMRG~\cite{Tenpy2018}, using the transfer matrix method in a specific sector, see Appendix~\ref{app:details} for details.

Fig.~\ref{fig:corr_length} shows the decay of the inverse correlation length as a function of the inverse bond dimension for three fillings close to the critical point $n_c=2/3$ inside the spin-gap phase. At $n\approx 0.583$ all correlation lengths diverge as bond dimension increases, which implies that the spin gap is very small. Furthermore, density correlation length is bigger than the pairing correlation length, therefore, the leading instability is charge density wave. As we increase the filling, the pairing correlation length exceeds the charge correlation length $\xi_p>\xi_{ns}$, which means that the charge Luttinger parameter $K_{c}>1$ and superconducting fluctuations dominate. 

Fig.~\ref{fig:corr_length}(c) shows the correlation lengths at $n\approx 0.633$ deep in the spin gap phase in proximity to the QCP. The spin ($\xi_s$) and charge ($\xi_c$) correlation lengths saturate, since the corresponding operators are short-ranged. We note, that the resulting spin correlation length  $\xi^\infty_s\approx 54$, where $\xi^\infty$ is the correlation length obtained from extrapolation to $\chi \rightarrow\infty$, see dashed lines of Fig.~\ref{fig:corr_length}(c). This means that one has to go to a very large system sizes to observe the spin gap in the finite DMRG methods.
In contrast, the density ($\xi_{ns}$) and pairing ($\xi_p$) correlation lengths diverge, and pairing correlation length is the biggest one: $\xi^{\infty}_p/\xi^{\infty}_c \approx4.1$ and $\xi^{\infty}_p/\xi^{\infty}_{ns} \approx2.1$. This clear hierarchy of correlation lengths provides strong evidence that this region of the phase diagram is a superconducting Luther-Emery liquid.

\begin{figure}[b!]
\begin{minipage}[h]{1\linewidth}
  \center{\includegraphics[width=0.75\linewidth]{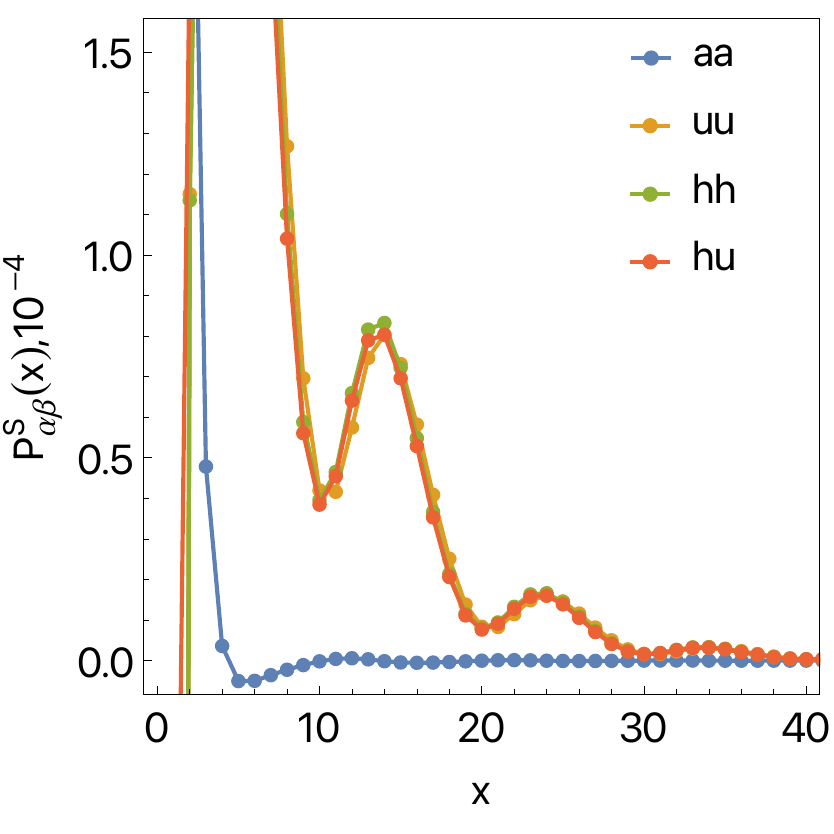}}
        \caption{The spin-singlet correlation function $P^S_{\alpha \beta}(x)$ at filling $n\approx0.63$ close to the critical point. The bond $a$ is between the $d$ and $p_x$ site. The bond $h$ is between $p_x$ and $p_y$ site inside the unit cell, while the bond $u$ is rotatted $90^\circ$ with respect to bond $h$. Calculations were performed using iDMRG on a unit cell $L_x=10$ with $N_e=38$ electrons, hubbard interaction $U=16$ and bond dimension $\chi=4000$. 
        \label{fig:supercond}}
    \end{minipage}
\end{figure}

Finally, we study the real space behavior of the pairing correlation function. The spin-singlet pair operator between sites $i$ and $j$ is defined as $\Delta_{ij}=c_{i\uparrow}c_{j\downarrow}-c_{i\downarrow}c_{j\uparrow}$ and the pairing correlation function is $P^{S}_{\alpha \beta}(x)=\langle \Delta^{\dag}_{\alpha}(x) \Delta^{}_{\beta}(0) \rangle  $, where $\alpha$ and $\beta$ define the bond. Fig.~\ref{fig:supercond} shows the decay of the pairing correlation function on different bonds.  We find that the strongest superconducting correlations are on the $h$ and $u$ bonds which link $p_x$ and $p_y$ sites, see Fig.~\ref{fig:geometry} for an illustration of the geometry. Furthermore, $P^S_{hh}(x)=P^S_{uu}(x)=P^S_{hu}(x)$ suggesting an $s_{xy}$-wave symmetry. We also observed that the spin-triplet operator $\Delta^T_{ij}=c_{i\uparrow}c_{j\downarrow}+c_{i\downarrow}c_{j\uparrow}$ and the corresponding triplet correlation function decays much faster than the singlet one for all relevant bonds, making $p$-wave pairing an unlikely scenario.

We note, that near a ferromagnetic QCP, the pairing interaction is usually attractive in the p-wave channel, while near an  antiferromagnetic QCP with a single Fermi surface the pairing interaction is attractive in the d-channel \cite{Abanov2020}. However, the Lieb lattice has ferrimagnetic order with both types of correlations present; therefore, the dominant superconducting channel could differ from the earlier established results. Furthermore, quasi-2D geometry could promote nematic instability, which is usually dominant in all channels. Therefore, more work is needed to establish the true character and the leading pairing channel near the magnetic QCP.

\section{conclusion}
\label{sec:conclusion}
In the present work we studied a Hubbard model on a two-leg Lieb lattice as a function of Hubbard interaction $U$ and electron filling $n$. At half-filling, the system develops a ferrimagnetic order, consistent with Lieb's original prediction. Away from half-filling, the ground state is metallic with the total spin $S^2_{tot}$ different from zero. 

The ferromagnetic order eventually disappears at $n<2/3$ giving rise to a Luttinger liquid. Its properties are generally explained from the weak-coupling limit: in particular it is demonstrated that such Liquid has one charge and one spin mode. Furthermore, at commensurate fillings $n=1/6$ and $n=1/2$ the charge mode becomes gapped, while the spin mode remains gapless. Note that such behavior is consistent with Lieb–Schultz–Mattis \cite{LSM}, since the extended unit cell has one or three electrons and thus a fully symmetric, nondegenerate, gapped state is forbidden. At $n=1/3$ both spin and charge modes become gapped and the resulting state is an insulator.

Interestingly, in the narrow doping range around $n_c=2/3$ we find a Luther Emery phase with a finite spin gap. Using iDMRG we studied the correlation lengths in this phase and identified superconductivity as the leading instability. Furthermore, by looking at pairing correlations in real space, we revealed that they are strongest on the bonds between $p_x$ and $p_y$ sites, and the corresponding superconducting order parameter has $s_{xy}$ symmetry.

We believe this work provides an important step toward understanding the full two-dimensional phase diagram of the Hubbard model on the Lieb lattice. In particular, the experimental realization of this model in optical lattices offers a controlled platform for testing our predictions. Furthermore, recent advances in Neural Quantum States \cite{carleo2017,Lange2024} open promising avenues for accessing the full two-dimensional limit. Finally, developing an analytical understanding of the emergent superconductivity near the ferrimagnetic critical point, as well as clarifying the specific role of flat bands, remains an important direction for future research.

\begin{acknowledgements}
We are grateful to Anant Kale, Yahui Zhang, Clemens Kuhlenkamp, Andrey Chubukov and Nishchhal Verma for useful discussions. This research was supported by
the U.S. National Science Foundation grant No. DMR 2245246 and by the Simons Collaboration on Ultra-Quantum Matter which is a grant from the Simons Foundation (651440, S.S.).

\end{acknowledgements}

\appendix

\section{DMRG details}
\label{app:details}
The DMRG calculation were performed using TeNPy library (version 0.10.0)~\cite{Tenpy2018}. The bond dimension $\chi$ varied from $500$ to $3000$ for finite DMRG and  from $500$ to $8000$ for iDMRG with the number of sweeps $n_{\text{sweeps}}=20$. The maximum discarded weight was below $10^{-5}$ and the energy converged to $10^{-8}$ between the sweeps.

The correlation lengths were computed using iDMRG from the eigenvalue spectrum of the transfer matrix, resolved in fixed symmetry sectors labeled by $(\delta Q, \delta S_z)$.
For instance, the charge correlation length ($\xi_c$) is extracted from the sector $(\delta Q, \delta S_z) = (1,1)$, which corresponds to operators carrying one unit of charge and spin projection $S_z = 1$, such as $c_\uparrow(x)$.
Table~\ref{table:corr} lists the symmetry sectors used to extract the various correlation lengths, together with representative operators associated with each sector.

As finite bond dimension $\chi$ introduces an effective infrared cutoff, the extracted correlation length $\xi_\chi$ will always be finite in any numerical calculation. In the Luttinger liquid phase, all correlation length diverge $\xi \propto \chi^\kappa$ as the bond dimension increases. Inside the spin gap phase, the charge and spin correlation lengths remain finite, while the density and pairing correlation lengths still diverge, the ratio between them determines whether the system is unstable towards a superconductor or a charge density wave.

\begin{table}[h]
\centering
\begin{tabular}{ccc}
\hline
sector $(\delta Q, \delta S_z)$ & Correlation length & Operator \\ 
\hline
$(0,0)$ & $\xi_{ns}$ & $N(x)$ \\

$(0,2)$ & $\xi_{s}$ & $S^+(x)$ \\

$(1,1)$ & $\xi_{c}$ & $c_{\uparrow}(x)$ \\

$(2,0)$ & $\xi_{p}$ & $\Delta_S(x)$ \\

$(2,2)$ & $\xi_{t}$ & $\Delta_T(x)$ \\
\hline
\end{tabular}
\caption{\label{table:corr}Charge sectors, corresponding correlation lengths, and operators.}
\end{table}

The central charge was found via iDMRG~\cite{Tenpy2018} using the entanglement entropy of a half-infinite chain $S=\frac{c}{6} \ln[\xi(\chi)]$, where $\xi(\chi)$ is the largest correlation length at a  fixed bond dimension $\chi$~\cite{Pollmann2009,Calabrese2004}.

\date{\today}

\bibliography{main}
\end{document}